\documentclass[12pt,a4paper]{article}
\usepackage{amssymb}
\usepackage{amsfonts}
\usepackage[english]{babel}
\usepackage{graphicx}
\usepackage{appendix}
\newcommand{\ba}{\begin{array}}
\newcommand{\ea}{\end{array}}
\newcommand{\pa}{\partial}

\newcommand{\no}{\nonumber}

\newcommand{\be}{\begin{equation}}
\newcommand{\ee}{\end{equation}}
\newcommand{\bea}{\begin{eqnarray}}
\newcommand{\eea}{\end{eqnarray}}
\newcommand{\beaa}{\begin{eqnarray*}}
\newcommand{\eeaa}{\end{eqnarray*}}

\begin{document}

\title
{ The resonant structure of Kink-Solitons in the Modified KP Equation}
\author{
 Jen-Hsu Chang \\Department of Computer Science and Information Engineering, \\
 National Defense University, \\
 Tauyuan City, Taiwan, 33551 }

\date{}

\maketitle
\begin{abstract}
Using the Wronskian representation of $\tau$-function, one can investigate the  resonant structure of kink-soliton and line-soliton of the modified KP equation. It is found that the resonant structure  of  the the soliton graph is obtained by superimposition of the two corresponding soliton graphs of the two Le-Diagrams given an irreducible Schubert cell in a  totally non-negative Grassmannian  $Gr(N,M)_{ \geq 0}$. Several examples are given.

Keywords: Wronskian, Grassmannian, Le-Diagram, Kink-Soliton

\end{abstract}

\newpage

\section{Introduction}
Recently, the resonant theory of line solitons has attracted much attention, especially in the KP-(II) theory \cite{bc,ko,ko1,ko2, ko3, ko5, ko6} and Novikov-Veselov (NV) equation \cite{jh1, jh2}. To study the resonant  theory of integrable models, we use the Wronskian representation of the $\tau$-function for the solution structure. Given an $ N \times M $-matrix ( $ N < M $), one can describe the $\tau$-function as a linear combination of exponential functions by Bitnet-Cauchy formula  and their coefficients have to satisfy the Plucker relations. It is required that all these  coefficients have to be  non-negative to get non-singular solutions. Therefore, one introduces the totally non-negative Grassmannian to the resonant  theory. Due to the success of resonant  theory in the KP-(II) equation, one can also investigate the solitonic resonance  of the Modified KP-(II) (MKP-(II)) equation, especially the resonant  theory of kink-soliton and line-soliton. It is noticed that the solution of MKP-(II) equation is associated with  the quotient of $\tau$-functions by the Wronskian representations, and then the parameters of the phases are also non-negative to obtain non-singular solutions (see below). Also, there is the Miura transformation between the KP-(II) equation and MKP-(II) equation.  \\
\indent  The MKP-(II) equation  is defined by \cite{dk, kn, kd, jm}
 \bea
-4v_{t}+v_{xxx}-6v^2v_{x}+6v_{x}\pa_x^{-1}v_y+3\pa_x^{-1}v_{yy}=0. \label{mkp}
 \eea
The equation (\ref{mkp}) was introduced in \cite{kn} within the framework of gauge-invariant description of the KP equation. In \cite{jm}, it appeared as the first member of modified KP hierarchy using the $\tau$-function theory. In \cite{kd}, the inverse-scattering-transformation  method  is used to get the exact solutions for MKP equation (\ref{mkp}), including rational solutions (lumps), line solitons and breathers. \\
\indent Letting 
\be v(x,y,t)= \pa_x \ln (P(x,y,t)/Q(x,y,t)), \label{tr}
\ee
we have the Hirota bi-linear equation \cite{hi, jm}
\bea
&& (D_y-D_x^2)P \circ Q =0  \label{h1} \\
&& (-4D_t +D_x^3+3 D_x D_y)P \circ Q =0  \label{h2}, 
\eea
where the bi-linear operators $D_x^m$ and $D_y^n$ are defined by 
\[D_x^m D_y^n P \circ Q = (\pa_x-\pa_{x^{'}})^{m} (\pa_y-\pa_{y^{'}})^{n} P(x,y) Q(x^{'},y^{'}).\]
To construct the solutions of these Hirota equations (\ref{h1}) and(\ref{h2}), one defines the determinant 
\be
\tau_N^{(n)}= det 
\left[\ba{cccc} g_1^{(n)}  & g_1^{(n+1)} & \cdots   &  g_1^{(n+N-1)}    \\
 g_2^{(n)} &  g_2^{(n+1)} & \cdots   &   g_2^{(n+N-1)} \\
\vdots  & \vdots & \vdots & \vdots    \\
 g_N^{(n)}& g_N^{(n+1)} & \cdots  &   g_N^{(n+N-1)}   \ea \right],    \ee
where the elements in the above determinant are defined by ($ i=1,2,3 \cdots, N $ )
\be  
\frac{\pa g_i}{\pa x_m}  = \frac{\pa^m g_i}{ \pa x^m} , \quad x_1=x, \quad x_2=y, \quad x_3=t,  \label{lin} 
\ee
and $g_i^{(n)}$ means the n-th order derivative with respect to $x$, $n=0,1,2,3, \cdots$. Also, we can write $\tau_N^{(n)}$ as a Wronskian, i.e., 
\[ \tau_N^{(n)}=Wr( g_1^{(n)}, g_2^{(n)}, g_3^{(n)}, \cdots, g_N^{(n)} ). \]
It is shown that \cite{hi}
\be
P=\tau_N^{(1)}, \quad Q=\tau_N^{(0)}, \quad or \quad  v(x,y,t)= \pa_x \ln \frac{\tau_N^{(1)}}{\tau_N^{(0)}} , \label{h3}
\ee 
will be  solutions of (\ref{h1}) and (\ref{h2}) for $N=1,2, \cdots $. \\
\indent We remark that after the Miura transformation \cite{kd}, using the Hirota equation (\ref{h1}), we have 
\bea 
u= - \pa_x^{-1}v_y-v_x-v^2=2\pa_{xx} \ln \tau_N^{(0)}, \label{mi}
\eea 
and then one can obtain the Hirota equation by (\ref{h2}) 
\[(-4D_t D_x +D_x^4+3  D_y^3) \tau_N^{(0)} \circ  \tau_N^{(0)}=0, \]
or the KP-(II) equation 
\bea 
-4u_t+u_{xxx}+6uu_x+\pa_x^{-1}3u_{yy}=0. \label{kp}
\eea
\indent Next, we construct the resonant solutions of MKP-(II) equation 
using the totally non-negative Grassmannian (TNNG) in KP-(II) theory \cite{  ko3, ko6}. Here one considers a finite dimensional solution
\beaa 
g_i (x,y,t)  &=& \sum_{j=1}^M k_{ij} H_j (x,y,t), \quad i=1,2, \cdots N < M, \\
H_j (x,y,t) &= & e^{\theta_j}, \quad  \theta_j= c_j x+c_j^2 y+ c_j^3 t + \xi_j, \quad j=1,2, \cdots M 
\eeaa 
where $c_j $ and $\xi_j$ are real parameters. For simplicity, we take $\xi_j=0$ in this article. Each $H_j(x,y,t)$ satisfies the equations (\ref{lin}). Then each resonant solution of MKP-(II) equation can be parametrized  by a full rank matrix 
\be K= \left[\ba{cccc} k_{11}  & k_{12}  & \cdots   &  k_{1M}   \\
 k_{21} &  k_{22} & \cdots   &   k_{2M} \\
\vdots  & \vdots & \vdots & \vdots    \\
 k_{N1} & k_{N2}  & \cdots  &   k_{NM}    \ea \right] \in M_{N \times M} (\textbf{R}). \label{km} \ee  
Using the Binet-Cauchy formula, the $\tau$-function $\tau_N^{(0)}$ can be written as 
\bea 
\tau_N^{(0)} &=& = \tau_K^{(0)}=\tau_N^{(0)} = Wr (g_1, g_2, \cdots, g_N)=det 
\left[\ba{cccc} g_1 & g_1^{'} & \cdots   &  g_1^{(N-1)}    \\
 g_2 &  g_2^{'} & \cdots   &   g_2^{(N-1)} \\
\vdots  & \vdots & \vdots & \vdots    \\
 g_N & g_N^{'} & \cdots  &   g_N^{(N-1)}   \ea \right]  \no \\
&=& det \left [\left(\ba{cccc} k_{11}  & k_{12}  & \cdots   &  k_{1M}   \\
 k_{21} &  k_{22} & \cdots   &   k_{2M} \\
\vdots  & \vdots & \vdots & \vdots    \\
 k_{N1} & k_{N2}  & \cdots  &   k_{NM}    \ea \right) \left(\ba{cccc} H_1  & c_1 H_1 & \cdots   &  c_1^{N-1}H_1   \\
 H_2 & c_2 H_2 & \cdots   &   c_2^{N-1} H_2 \\
\vdots  & \vdots & \vdots & \vdots    \\
 H_M & c_M H_M  & \cdots  &    c_M^{N-1} H_M   \ea \right)  \right] \no \\
&=& \sum_J \Delta_J (K) H_J (x,y,t), \label{ta}
\eea 
where $\Delta_J (K)$ is the  $ N \times N $ minor for the columns  with the index set $J=\{ j_1, j_2, j_3, \cdots, j_N\} $, and $H_J$ is the Wronskian 
\be  H_J= Wr (H_{j_1},H_{j_2}, H_{j_3}, \cdots, H_{j_N} )= \prod_{m< l} (c_{j_l}-c_{j_m}) H_{j_1} H_{j_2}H_{j_3} \cdots H_{j_N}. \label{va} \ee
We notice that the coefficients $ \Delta_J (K)$ of $\tau_K^{(0)}$  have to satisfy the Plucker relations. \\
\indent Similarly, 
\bea 
\tau_K^{(1)} &=& \tau_N^{(1)} = Wr (g_1^{'}, g_2^{'}, \cdots, g_N^{'}) =det 
\left[\ba{cccc} g_1^{'} & g_1^{''} & \cdots   &  g_1^{(N)}    \\
 g_2^{'} &  g_2^{''} & \cdots   &   g_2^{(N)} \\
\vdots  & \vdots & \vdots & \vdots    \\
 g_N^{'} & g_N^{''} & \cdots  &   g_N^{(N)}   \ea \right]  \no \\
&=& det \left [\left(\ba{cccc} c_1 k_{11}  & c_2 k_{12}  & \cdots   & c_M k_{1M}   \\
c_1  k_{21} &  c_2 k_{22} & \cdots   &   c_M k_{2M} \\
\vdots  & \vdots & \vdots & \vdots    \\
 c_1 k_{N1} & c_2 k_{N2}  & \cdots  &  c_M  k_{NM}    \ea \right) \left(\ba{cccc} H_1  & c_1 H_1 & \cdots   &  c_1^{N-1}H_1   \\
 H_2 & c_2 H_2 & \cdots   &   c_2^{N-1} H_2 \\
\vdots  & \vdots & \vdots & \vdots    \\
 H_M & c_M H_M  & \cdots  &    c_M^{N-1} H_M   \ea \right)  \right] \no \\
&=& \sum_J \Delta_J (K) c_{j_1}  c_{j_2}  c_{j_3} \cdots    c_{j_N} H_J (x,y,t).  \label{taa}
\eea 
We notice here that the functions $\tau_N^{(0)}$ and  $\tau_N^{(1)} $ in (\ref{ta}) and (\ref{taa}) are  invariant under the transformations $  K \to  \Gamma K $ in (\ref{km}),  $ \Gamma$ being a constant $N \times N$ matrix, due to the solution structure (\ref{h3}). As a result, the K-matrix in (\ref{km}) can be chosen as a reduced row echelon form (RREF). Also, we can define a K-matrix to be irreducible in RREF if \\
(1) in each column, there is at least one non-zero element; \\
(2) in each row, there is at least one more non-zero element in addition to the pivot. \\
\indent On the other hand, using the Plucker embedding \cite{ko1, ko6},  one can consider the real Grassmannian $Gr(N,M) \cong GL_N(\textit{R}) \backslash M_{N \times M} (\textit{R})$. The Schubert decomposition  of  $Gr(N,M)$ is 
\[ Gr(N,M)= \bigcup_I \Omega_I,  \]
where $\Omega_I$ (a Schubert cell)  in RREF is defined by the set of all matrices whose pivots are given by $I= \{i_1, i_2, i_3, \cdots, i_N\} $.  For each Schubert cell $\Omega_I$, one introduces the Young diagram to express the index set $I$ for an alternative parametrization. To obtain non-singular solutions of MKP-(II), from (\ref{h3}),  (\ref{ta}) and  (\ref{taa}), it can be seen that  $ \Delta_J (K) \geq 0$ for all $J$, i.e.,  K is an element of  the TNNG $Gr(N,M)_{ \geq 0}$, i.e., the corresponding Plucker coordinates of each $\Omega_I$ are non-negative. And we assume the ordering in the $c$-parameters, 
\be  0 \leq c_1 < c_2 < c_3 < \cdots < c_M. \label{or} \ee  
In the KP-(II) equation, we have the following Classification Theorem  for the unbounded line solitons as $ y \to \pm \infty $ \\
{\bf Theorem \cite{bc}:}  Let $ \{e_1, e_2, e_3, \cdots, e_N \} $  be the pivot indices, and let $\{ f_1, f_2, \cdots, f_{M-N} \} $ be the non-pivot indices for an irreducible and totally non-negative K-matrix. Then the soliton solution associated with the K-matrix has \\
(a) $N$-line solitons  of $[e_n, j_n]$-type for $n=1,2, \cdots, N$ as $y \to \infty$ \\
(b) $(M-N)$-line solitons of $[i_m, f_m]$-type for $m=1,2, \cdots, M-N$ as $y \to -\infty$. \\
The set of those unbounded line solitons $[e_n, j_n]$ and  $[i_m, f_m]$ are expressed by an unique derangement (a permutation of $\{1,2,3,\cdots, M\}$ without any fixed point).  \\
\indent For an irreducible  $\Omega_I \in Gr(N,M)_{ \geq 0} $, we can find its associated Le-diagram through the  pipe dreams \cite{ko3, po}. A Le-Diagram is a Young diagram filling with ” + ” or ” $\circ $ ” in each box, which has the Le-property: there is no ”  $\circ $  ” which has ” + ” above it AND its left. Then we have \\
{\bf Theorem \cite{po}:}There is a bijection between the set of irreducible Le-Diagrams and the set of derangements.\\
Therefore one can represent an irreducible Schubert cell $\Omega_I \in Gr(N,M)_{ \geq 0}$  with a Le-Diagram, and conversely. Then the corresponding soliton graph is obtained from the Le-Diagram. Much more detail can be found in  \cite{ko3,ko6}.

This paper is organized as follows. In next section, one describes the resonant structure by the superimposition of soliton graphs of Le-Diagrams. In section 3,  several examples are given to illustrate the method in section 2. In section 4,  we conclude the paper with several remarks.

\section{ Resonant Structure }
In this section, one constructs the resonant structure. It is the superimposition of soliton graphs of the Le-Diagrams. For  basic  solutions of  interaction between line solitons and multi-kinks solitons, one refers to \cite{jh3}.  \\
\indent As in the KP-(II) case \cite{bc, ko}, from the form of $\tau$-function (\ref{ta}) and (\ref{taa}),  each line soliton is obtained by the balance between $H_J$ and $H_{J'}$ in the $xy$-plane and happens  only at the boundaries of the dominant regions. The line solitons  are  obtained in adjacent regions of $xy$-plane contain $N-1$ common  phases and differ by only one single phase.  Now, suppose $ H_{i, j_2, j_3, \cdots, j_N}$ and $ H_{j, j_2, j_3, \cdots, j_N} $ in (\ref {va}) are adjacent regions. Then one has the boundary, i.e., the  line soliton $ [i,j]$-soliton \cite{jh3} 
\be  v \approx  \frac{c_j-c_i}{2} (\tanh \frac{\Omega_j-\Omega_i+\ln \frac{c_j}{c_i} }{2}-\tanh \frac{\Omega_j-\Omega_i}{2}) \geq 0, \label{prg} \ee
where 
\[ \Omega_j=\theta_j+ \ln \vert \prod_{m=2}^N (c_j-c_{j_m})\vert , \quad \Omega_i=\theta_i+ \ln \vert \prod_{m=2}^N (c_i-c_{j_m})\vert . \]
Also, when $ \Omega_j-\Omega_i=-\frac{1}{2} \ln \frac{c_j}{c_i}$, this $ [i,j]$-soliton has maximal value $ (\sqrt{c_j}-\sqrt{c_i})^2$. When $c_1 \neq 0$,   the  line $ [i,j]$-soliton  graph 
(\ref{prg}) of MKP equation is similar to the KP equation. Similar to the case KP-(II) \cite{ko1}, it can be seen that 
the $[i,j]$-line soliton solution (\ref{prg}) has the wave vector 
\be \vec {\Sigma}_{[i,j]}=(c_j-c_i, c_j^2-c_i^2),  \label{tr} \ee 
and can be measured in the counterclockwise sense from the $y$-axis, i.e., 
\be \tan \Phi_{[i,j]}=\frac{c_j^2-c_i^2}{c_j-c_i}=c_i+c_j, \label{po} \ee
moreover, its velocity is given by 
\be \vec {V}_{[i,j]}= \frac{c_i^2+c_ic_j+c_j^2}{1+(c_i+c_j)^2} (1, c_i+c_j), \label{ve} \ee
and the frequency is given by
\be \Delta_{i,j}= c_j^3-c_i^3=(c_j-c_i)(c_i^2+c_ic_j+c_j^2).  \label{fe} \ee
Similar to the KP-(II) equation \cite{ko1}, three line solitons can interact to form a trivalent vertex and satisfy the resonant conditions for wave number and frequency by (\ref{tr}) and (\ref{fe}) ($i < m < j $)
\be  \vec{\Sigma}_{[i,j]}= \vec{\Sigma}_{[i,m]}+ \vec{\Sigma}_{[m,j]}, \quad   {\Delta}_{[i,j]} ={\Delta}_{[i,m]} + {\Delta}_{[m,j]}.   \label{re} \ee
On the other hand, a multi-kinks solution can be obtained by $c_1=0$ \cite{jh3}. In this case, each  $[1,j]$-line soliton in  (\ref{prg}) becomes kink front, i.e., 
\be  v \approx \frac{c_j}{2} (1-\tanh \frac{\Omega_j- \ln  \prod_{m=2}^N c_{1_m} }{2} ) \to  \left\{\ba {ll} c_j , &  x   \to -\infty ,  \\   0 ,  &  x  \to \infty, \ea \right.  \label{mk} \ee
and forms the boundary of the multi-kinks solution. The front of the multi-kinks solution of (\ref{mk}) is defined as
\be  \Omega_j- \ln  \prod_{m=2}^N c_{1_m} =0.  \label{bd} \ee
We notice that in the MKP equation line solitons can pierce fronts of the multi-kinks solutions to form  new boundaries by resonance  when $c_1=0$ \cite{jh3}, which is different from the KP equation. The resonant structure of  line solitons and  the multi-kinks solutions of the MKP equation is the main purpose for our investigation. \\
\indent To this end, one considers the equation (\ref{taa}). When $c_1=0$, it is equivalent to deleting the first column vector of $K$-matrix. Therefore, we have to consider the  TNNG $Gr(N,M-1)_{ \geq 0}$. Also from (\ref{h3}) it's known that 
\[ v=\pa_x \ln \tau_N^{(1)}- \pa_x \ln \tau_N^{(0)}. \]
Our main observation is that when $c_1=0$ the resonant structure of MKP equation is 
\be    Gr(N,M-1)_{ \geq 0}+ Gr(N,M)_{ \geq 0}, \label{led}   \ee
i.e, the soliton graph is the superimposition of the soliton graphs of the  Le-Diagrams of the corresponding totally non-negative Grassmannians $Gr(N,M-1)_{ \geq 0}$ and  $Gr(N,M)_{ \geq 0}$.  We remark that  $Gr(N,M-1)_{ \geq 0}$ here may be not irreducible and then the corresponding permutation is not of derangement due to deleting the first column vector of $K$-matrix, i.e., it has some  fixed points. Also, one can think  the Le-Diagram of the corresponding $Gr(N,M-1)_{ \geq 0}$ as the soliton graph for $c_1=0$ obtained from (\ref{mk}) and the resonance (\ref{re}) of line solitons  (\ref{prg}) piercing the front (\ref{bd}). \\
\indent Next, we describe how to construct the soliton graph given a K-matrix (irreducible in RREF) $ \in Gr(N,M)_{ \geq 0}$. Firstly, one  constructs the Le-Diagrams $\textsl{L}^{(M)}_{\pm}$ ( $\textsl{L}^{(M)}_{-}=\textsl{L} $ and $\textsl{L}^{(M)}_{+}=\textsl{L}^\ast  $ in \cite{ko3} )  corresponding to $Gr(N,M)_{ \geq 0}$  for $c_1 \neq 0$  when $t \to \pm \infty$. The soliton graph  is obtained similarly to the KP equation \cite{ko1, ko3}. Secondly, one constructs the soliton graphs for $c_1=0$  as follows when $t \to \pm \infty$. 

\begin{itemize}
\item Delete the first column of  the Le-Diagram $\textsl{L}^{(M)}_{+}$ to get the Le-Diagram $\textsl{L}^{(M-1)}_{+}$ corresponding to $Gr(N,M-1)_{ \geq 0}$ and its corresponding permutation of $ \{2, 3, 4, \cdots, M \}$ $ \pi_+[N,M-1] $ by pipe dreams.  
\item Then from the  corresponding permutation  $ \pi_{-}[N,M-1]= ( \pi_+[N,M-1])^{-1}$ one obtains the Le-Diagram $\textsl{L}^{(M-1)}_{-}$ also by pipe dreams.  
\item When $t \to \infty $,  the soliton graph is obtained by superimposition of the soliton graphs of the Le-Diagrams $\textsl{L}^{(M)}_{+}$ and $\textsl{L}^{(M-1)}_{+}$. 
\item When $t \to -\infty $,  the soliton graph is also obtained by  superimposition of the soliton graphs of the Le-Diagrams $\textsl{L}^{(M)}_{-}$ and $\textsl{L}^{(M-1)}_{-}$. 
\end{itemize}
One has two remarks here. \\
(1) The permutation $ \pi_+ [N,M-1]$ may be not a derangement, i.e., it has some fixed points,  as mentioned previously. Therefore, the permutation  $\pi_{-}[N,M-1]$ is not a derangement, either.  \\
(2) If K-matrix is an element of the TNNG, then every $N \times N$ sub-determinant is non-negative. As a result, the matrix obtained by deleting the first column of K-matrix is still an element of the TNNG.

\section{Examples}
In this section, we give several examples to illustrate the procedure described  in last section. For simplicity, we focus on the TNNG $Gr(2,4)$ \cite{ko1}. 
\begin{itemize} 
\item  One considers the $\tau$-function 
\[ \tau_5=H_1+a H_2+b H_3+c H_4 +d H_5,\]
where $a, b, c, d >0 $ and it corresponds the TNNG $Gr(1,5)$. Then
\[\tau_{5x}=k_1 H_1+k_2 a H_2+k_3 b H_3+k_4cH_4 +k_5 d H_5.\]
So we  have the solution of MKP-(II)  
\be  u=\pa_x \ln \frac{\tau_{5x}}{\tau_{5}}=\pa_x \ln \frac{ k_1 H_1+k_2 a H_2+k_3 b H_3+k_4cH_4 +k_5 d H_5    }{ H_1+a H_2+b H_3+c H_4 +d H_5      }. \label{l5} \ee
Please see the figure 1 for the soliton graphs and the corresponding Le-diagrams (or the permutations) when $t \to \pm \infty$. For $ t  \to \pm \infty $,  one obtains unbounded line solitons (from left to right):   
\begin{itemize}
\item for $ y >> 0$, [2,5], [1,5] 
\item for $ y << 0 $, [1,2] , [2,3], [3,4], [4,5] . 
\end{itemize} 
The kink-fronts are  $[1,2]$ and $ [1,5]$, and the line solitons  $[3,4]$,  $ [4,5]$ and $[3,5]$ form a resonant trivalent vertex (\ref{re}). Also, the line soliton $[2,5]$ pierce the front $[1,2]$ and then with the front $ [1,5]$ form a unbounded terrace of height $k_5=1$. For $ t << 0$,  the unbounded line solitons are the the same as $ t >> 0$; moreover, we see that $[1,2], [1,3], [1,4]$ and   $ [1,5] $ form the multi-kink fronts and look like a refraction of light. On the other hand, the kink-front $[2,4]$  is the resonance of line solitons $[3,4]$ and $[2,3]$, and then form the quadrilateral of height $k_4=0.7$ with the kink-front $[1,4]$ and the line solitons $[3,4]$ and $[4,5]$ piercing the fronts. \\
\indent When comparing with the  unbound  line solitons in KP-(II) equation, i.e., $\vert y \vert \to \infty $, we know that the line soliton $[2,5]$ is the only difference, i.e., there is no such unbounded kink front  in  KP-(II) equation . It is explained by the corresponding Le-Diagrams. The permutations corresponding to $ t \to \infty$ are  $\pi_{+}[1,5]=(23451)$ and $\pi_{+}[1,4]=(3452)$; moreover, the permutations corresponding to $ t \to - \infty$ are  $\pi_{-}[1,5]=(51234)=(\pi_{+}[1,5])^{-1}$ and $\pi_{-}[1,4]=(5234)=(\pi_{+}[1,4])^{-1}$. \\
\indent For general case, it is similar. One considers ( $M \geq 4$ )  
\[ \tau_M= H_1+a_2 H_2+a_3 H_3+\cdots +a_M  H_M,    \]
where $ a_2, a_3, a_4, \cdots, a_M >0 $ and it corresponds the TNNG $Gr(1,M)$. Then
\[\tau_{Mx}= k_1H_1+a_2k_2 H_2+a_3 k_3 H_3+\cdots +a_M k_M H_M.\]
Then we have the  solution of MKP-(II)  
\be  u=\pa_x \ln \frac{\tau_{Mx}}{\tau_{M}}=\pa_x \ln \frac{k_1H_1+a_2k_2 H_2+a_3 k_3 H_3+\cdots +a_M k_M H_M }{ H_1+a_2 H_2+a_3 H_3+\cdots +a_M  H_M  }. \label{ln} \ee
For $ t >> 0$, one obtains unbounded line solitons  (from left to right):   
\begin{itemize}
\item for $ y >> 0 , [2,M], [1,M] $ 
\item for $ y << 0 $, $ [1,2] ,  [2,3], [3,4], [4,5], \cdots, [M-1, M-2], \cdots [M-1,M] $. 
\end{itemize} 
But $[2,M]$ and $ [1,M]$ form a smaller unbounded terrace of height $k_M$. For $ t << 0$, similarly, the unbounded line solitons are the same as $ t >> 0$ and  $[1,2], [1,3], [1,4], \cdots, [1,M-1]$ and   $ [1,M] $ form the multi-kink fronts. Also, there are $(M-3)$ quadrilaterals of different heights coming from the resonance condition (\ref{re}), composed of line solitons piercing the fronts.  The Le-Diagrams look like the figure 1. The permutations corresponding to $ t \to  \infty$ are  $\pi_{+}[1,M]=(2345\cdots M 1)$ and $\pi_{+}[1,M-1]=(34 \cdots M 2)$;  moreover, the permutations corresponding to $ t \to - \infty$ are  $\pi_{-}[1,M]=(M 12 \cdots M-1)=(\pi_{+}[1,M])^{-1}$ and $\pi_{-}[1,M-1]=(M 23\cdots M-1)=(\pi_{+}[1,M-1])^{-1}$. \\ 
\indent One notices here that in \cite{br} there are also have the soliton graphs of interaction between kink-soliton and line soliton but their solution structure is different from the one (\ref{h3}) represented by the Wronskian representations  (\ref{ta}) and  (\ref{taa}).

\item The  Grassmannian of T-type has the RREF
\[ K_T=\left[\ba{cccc} 1 & 0 & -c & -d  \\ 0 & 1 & a & b  \ea
 \right], \]
where  $a, b, c$ and $d$ are positive numbers, and $ ad-bc >0$. It is of the most complicated TNNG in $Gr(2,4)$ and the total number of non-zero minors $\Delta_J$ is six. The corresponding $\tau$-functions are 
\beaa \tau_{T}^{(0)} &=&   (c_2-c_1)H_1 H_2 +a(c_3-c_1)H_3H_1 +b(c_4-c_1)H_1H_4 +c(c_3-c_2)H_2 H_3 \\ 
& + & d (c_4-c_2)H_2 H_4 + (ad-bc)(c_4-c_3)H_3 H_4,   \eeaa
and 
\beaa \tau_{T}^{(1)} &=&   (c_2-c_1)c_1 c_2 H_1 H_2 +a(c_3-c_1)c_1 c_3H_1H_3 +b(c_4-c_1)c_1 c_4 H_1H_4 \\ 
&+& c(c_3-c_2)c_2 c_3H_2 H_3 +  d (c_4-c_2)c_2 c_4 H_2 H_4 + (ad-bc)(c_4-c_3)c_3 c_4H_3 H_4.   \eeaa
Please see the figure 2 for the soliton graphs and the corresponding Le-diagrams (or the permutations) when $t \to \pm \infty$. For $ t \to \pm \infty $,  one obtains unbounded line solitons or fronts (from left to right):   
\begin{itemize}
\item for $ y >> 0$, [3,4], [2,4], [2,3], [1,3];  
\item for $ y << 0 $, [1,3] , [2,4]. 
\end{itemize} 
When $t \to \infty $, the kink-fronts are  $[1,2]$ , $[1,3]$ and $ [1,4]$, and the line solitons  $[3,4]$,  $ [2,4 ]$ and $[2,3]$ form a resonant trivalent vertex (\ref{re}). Also, the line solitons 
$[3,4]$ and $[2,3]$ pierce the fronts $[1,4]$ and $ [1,3]$, and  then with the line soliton $ [2,4]$ and front $ [1,3]$ form two unbounded terraces of heights $k_4=3.5$ and $k_3=2$, respectively. For $ t << 0$, we also see that $[1,2], [1,3]$  and  $ [1,4] $ form the multi-kink fronts. On the other hand, the kink-front $[3,4]$  is the resonance of kink fronts  $[2,4]$ and $[2,3]$, and then form the terrace  of height $k_4=3.5$. We notice the front $[1,2]$ has the height $k_2=1$. \\
\indent When comparing with the  unbound  line solitons in KP-(II) equation, i.e., $ y  \to \pm \infty $, one knows that the fronts  $[2,3]$ and $[3,4]$ in figure 2 are different, i.e., there are no such unbounded kink fronts  in  KP-(II) equation. It can be read off  by the corresponding Le-Diagrams. The permutations corresponding to $ t \to \infty$ are  $\pi_{+}[2,4] (K_T)=(3412)$ and $\pi_{+}[2,3](K_T)=(342)$; moreover, the permutations corresponding to $ t \to - \infty$ are  $\pi_{-}[2,4] (K_T)=(3412)=(\pi_{+}[2,4](K_T))^{-1}=(3412)^{-1}$ and $\pi_{-}[2,3](K_T)=(423)=(\pi_{+}[2,3](K_T))^{-1}=(342)^{-1}$. \\
In \cite{lp}, the authors constructed the solutions of  MKP equation using the Kaup-Newell hierarchy, and then  the resonance line solitons different from the Wronskian representation (\ref{h3}) were obtained by the Hirota method. Their Hirota constraint plays the similar role as the T-type.  The relations between these T-type solitons of the MKP equation are not clear and need further investigation. 
\item The  Grassmannian of reduced T-type has the same RREF as $K_T$ but $ ad-bc =0$. Then the corresponding $\tau$-functions are 
\beaa \tau_{T^r}^{(0)} &=&   (c_2-c_1)H_1 H_2 +a(c_3-c_1)H_3H_1 +b(c_4-c_1)H_1H_4 +c(c_3-c_2)H_2 H_3 \\ 
& + & d (c_4-c_2)H_2 H_4,   \eeaa
and 
\beaa \tau_{T^r}^{(1)} &=&   (c_2-c_1)c_1 c_2 H_1 H_2 +a(c_3-c_1)c_1 c_3H_1H_3 +b(c_4-c_1)c_1 c_4 H_1H_4 \\ 
&+& c(c_3-c_2)c_2 c_3H_2 H_3 +  d (c_4-c_2)c_2 c_4 H_2 H_4.   \eeaa

Please see the figure 3 for the soliton graphs and the corresponding Le-diagrams when $t \to \pm \infty$. For $ t \to \pm \infty $,  one obtains unbounded line solitons or fronts (from left to right):   
\begin{itemize}
\item for $ y >> 0$, [3,4], [2,4], [1,2];  
\item for $ y << 0 $, [1,3] , [3,4]
\end{itemize} 
When $t \to \infty $, the kink-fronts are  $[1,2]$ , $[1,3]$ and $ [1,4]$, and the fronts  $[1,2]$,  $ [1,4 ]$ and $[2,4]$ form a resonant trivalent vertex (\ref{re}). Also, the line soliton 
$[3,4]$ pierces the front $[1,4]$, and  then with the front  $ [2,4]$ forms a unbounded terrace of heights $k_4=3.5$. For $ t << 0$, the fronts  $[1,2]$,  $ [1,3 ]$ and $[2,3]$ form a resonant trivalent vertex, and then we see the kink-front $[2,4]$  is the resonance of kink front  $[2,3]$ and the line soliton $[3,4]$ piercing the front $ [1,2] $. We notice the interaction between the front $ [1,2] $ and the line soliton $[3,4]$ is of O-type \cite{jh3}. \\
\indent Comparing with the  unbound  line solitons in KP-(II) equation for  $ y  \to \pm \infty $, we know that only the front $[3,4]$ in figure 3 is different. It also can  be read off  by the corresponding Le-Diagrams. The permutations corresponding to $ t \to \infty$ are  $\pi_{+}[2,4] (K_T^r)=(3142)$ and $\pi_{+}[2,3](K_T^r)=(243)$; moreover, the permutations corresponding to $ t \to - \infty$ are  $\pi_{-}[2,4] (K_T^r)=(2413)=(\pi_{+}[2,4](K_T^r))^{-1}=(3142)^{-1}$ and $\pi_{-}[2,3](K_T^r)=(243)=(\pi_{+}[2,3](K_T^r))^{-1}=(243)^{-1}$. We notice here that the permutations $\pi_{+}[2,3](K_T^r)=(243)$ and $ \pi_{-}[2,3](K_T^r)=(243) $ both have the fixed point $"2"$. 

\item The  Grassmannian of Mach-type has the RREF
\[ K_M=\left[\ba{cccc} 1 & a & 0 & -c  \\ 0 & 0 & 1 & b  \ea
 \right], \]
where  $a, b$ and $ c$ are positive numbers.  The corresponding $\tau$-functions are 
\beaa \tau_M^{(0)} &=&   (c_3-c_1)H_3H_1 +b(c_4-c_1)H_1H_4 +a(c_3-c_2)H_2 H_3 \\ 
& + & ab (c_4-c_2)H_2 H_4 + c(c_4-c_3)H_3 H_4,   \eeaa
and 
\beaa \tau_M^{(1)} &=& (c_3-c_1)c_1 c_3H_1H_3 +b(c_4-c_1)c_1 c_4 H_1H_4 \\ 
&+& a(c_3-c_2)c_2 c_3 H_2 H_3 +  ab  (c_4-c_2)c_2 c_4 H_2 H_4 + c(c_4-c_3)c_3 c_4H_3 H_4.   \eeaa

Please see the figure 4 for the soliton graphs and the corresponding Le-diagrams  when $t \to \pm \infty$. For $ t \to \pm \infty $,  one obtains unbounded line solitons or fronts (from left to right):   
\begin{itemize}
\item for $ y >> 0$, [3,4], [2,3], [1,3];  
\item for $ y << 0 $, [1,2] , [2,4]. 
\end{itemize} 
When $t \to \infty $, the kink-fronts are  $[1,2]$ and  $[1,3]$, and the line solitons  $[3,4]$,  $ [2,4 ]$ and $[2,3]$ form a resonant trivalent vertex (\ref{re}). Also, the line soliton
$[2,3]$ pierces the front  $ [1,3]$, and  then with the  front $ [1,3]$ forms a unbounded terrace of height  $k_3=1.4$; moreover, the interaction between the front $ [1,2] $ and the line soliton $[3,4]$ is of O-type \cite{jh3}. \\   For $ t << 0$, we  see that $[1,2]$, $ [1,3]$  and  $ [1,4] $ form the multi-kink fronts. On the other hand, the kink-front $[1,4]$  is the resonance of kink front  $[1,2]$ and the line soliton $[2,4]$, and then form the bounded terrace  of height $k_4=2.5$ which has the highest amplitude. We notice the front $[1,2]$ has the height $k_2=1$. \\
\indent Comparing with the  unbound  line solitons in KP-(II) equation for  $ y  \to \pm \infty $, one knows that the front  $[2,3]$ in figure 4 is  different. It can be read off  by the corresponding Le-Diagrams. The permutations corresponding to $ t \to \infty$ are  $\pi_{+}[2,4] (K_M)=(2413)$ and $\pi_{+}[2,3](K_M)=(423)$; moreover, the permutations corresponding to $ t \to - \infty$ are  $\pi_{-}[2,4] (K_M)=(3142)=(\pi_{+}[2,4](K_T))^{-1}=(2413)^{-1}$ and $\pi_{-}[2,3](K_T)=(342)=(\pi_{+}[2,3](K_M))^{-1}=(423)^{-1}$. \\

\end{itemize}

\section{Concluding Remarks}
In this article, using the Wronskian structure of $\tau$-functions, we investigate the resonant structure of kink-solitons and line solitons by the Le-Diagrams given an irreducible Schubert cell $\Omega_I \in Gr(N,M)_{ \geq 0}$. It turns out that the soliton graph  comes from  the superimposition of the soliton graphs of the  Le-Diagrams of the corresponding totally non-negative Grassmannians $Gr(N,M-1)_{ \geq 0}$ and  $Gr(N,M)_{ \geq 0}$. To illustrate the construction, one  gives several examples focusing on  $Gr(2,4)_{ \geq 0}$  for simplicity. Also, we make a comparison with the KP equation when $c_1=0$, and the main difference can be described by the Le-Diagram of  $Gr(N,M-1)_{ \geq 0}$. Some line solitons may pierce fronts to form bounded terraces and unbounded terraces  with the kink solitons or by resonance with the kink solitons.   \\
\indent    In addition, the Mach type kink soliton $K_M$ could be interesting \cite{jh1, ko1, ko5}. It is known that the Mach stem wave,  the bounded terrace  of height $k_4$ in figure 4, has the highest amplitude and it is the resonance of kink soliton $[1,3]$ (incidence   wave)  and line soliton  $[3,4]$ (reflected wave) \cite{ko5}. There is a critical angle  between O-type soliton and Mach type soliton  in the KP-(II) equation \cite{ko5} or the Veselov-Novikov equation \cite{jh1}. The critical angle depends on the initial angle of V-shape. It could be noteworthy to find the critical angle and its relation with the amplitude. Besides, given a portion of a permutation or a partial chord diagram as the initial value problem, one has to consider a completion of this partial chord diagram for convergence \cite{ko1}.  Completion means adding other solitons (kink solitons) or chords completes the initial diagram.  The completion  may not be unique. In \cite{ko1}, it is proposed a concept of minimal completion in the sense that the completed diagram has the minimal length of the chords  and the corresponding TNNG cell has the minimal dimension. It is very interesting to investigate the minimal completion for the line solitons or kink solitons in the MKP equation even though there is the Miura transformation between KP equation and MKP equation. Also, the condition (\ref{or}) can't be  used to get the solitons for the  Modified KdV (MKdV) equation \cite{koo}. On the other hand, by the complex conjugate pairs  of $c_i$, breather solutions in Wronskian form for MKdV equation are also constructed in \cite {sp, zz}, and there exist breather solutions for MKP equation  \cite{kd, kon}.  One can hope the complex Grassmannian theory could be generalized to the breather solutions using the correspondence between the Wronskian and Grammian forms of $\tau$-function \cite{ck,hw}. Finally, we notice  that the two-solitons solution of Kaup-Newell equation, a reduction of MKP equation, corresponds to the T-type soliton of MKP equation \cite{lp}.  In \cite{os}, the constrained MKP flows is  constructed and then one has the the reduction of MKP hierarchy. The solitons solution can be constructed using the Hirota bilinear method but may be not in Wronskian representation. On the other hand, the T-type soliton is generalized in \cite{ko2} (self-dual $\tau$-function). The relations between the reduction in \cite{os}  and the corresponding Schubert cell of self-dual $\tau$-function in \cite{ko2} would be interesting. These issues will be published elsewhere.

\subsection*{Acknowledgments}
The author thanks Prof. Y. Kodama for his very valuable suggestions. This work is supported  by the Ministry of Science and Technology of Taiwan under Grant No.: 106-2115-M-606-001.

\begin{figure}[b]
	\includegraphics[width=1.2\textwidth]{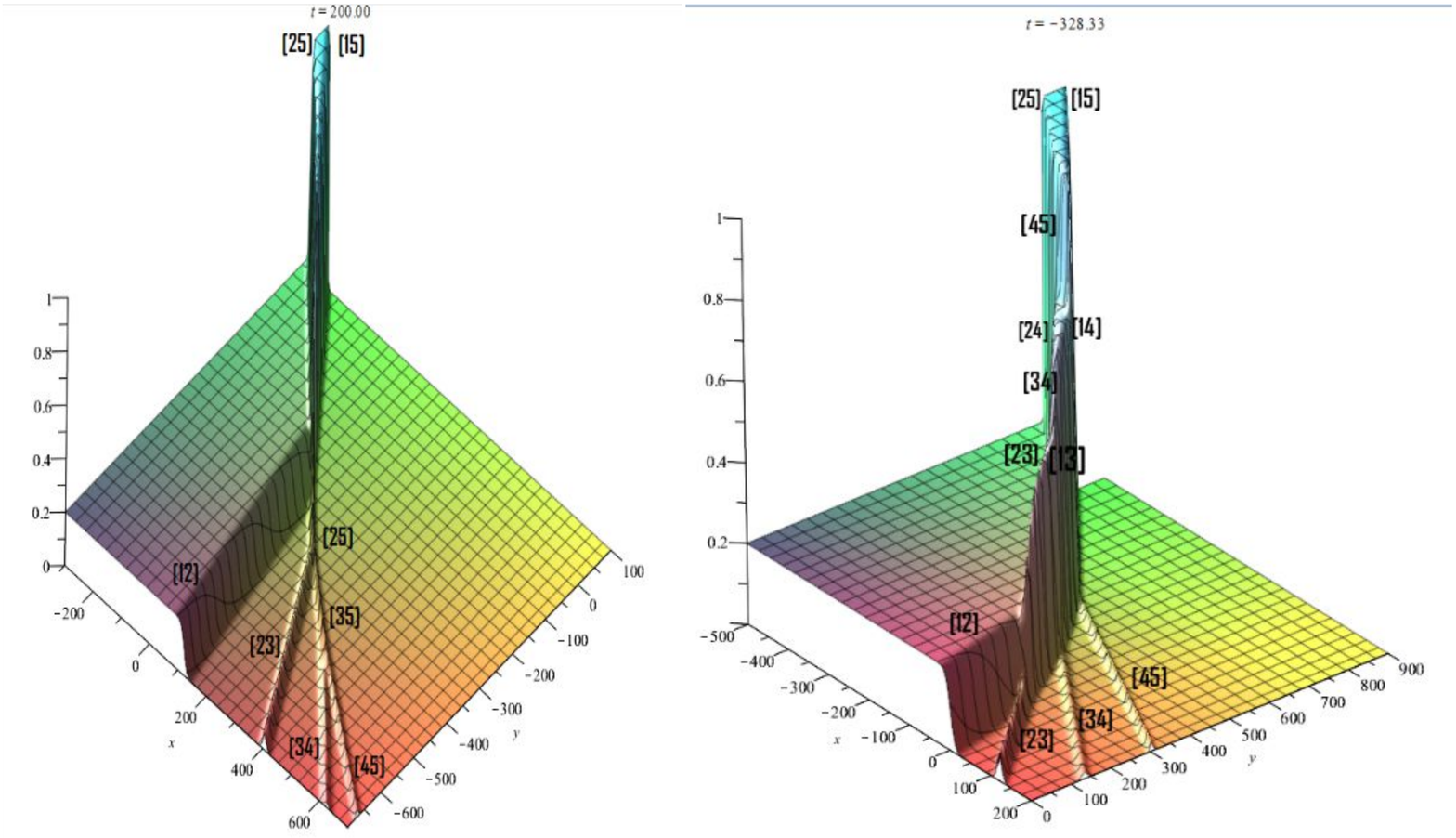}
	\caption{{ Refraction ($ k_1=0, k_2=0.2, k_3=0.5, k_4=0.7, k_5=1, a=10, b=40, c=20, d=30 $ ) }}
\end{figure}

\begin{figure}[b]
		\includegraphics[width=1\textwidth]{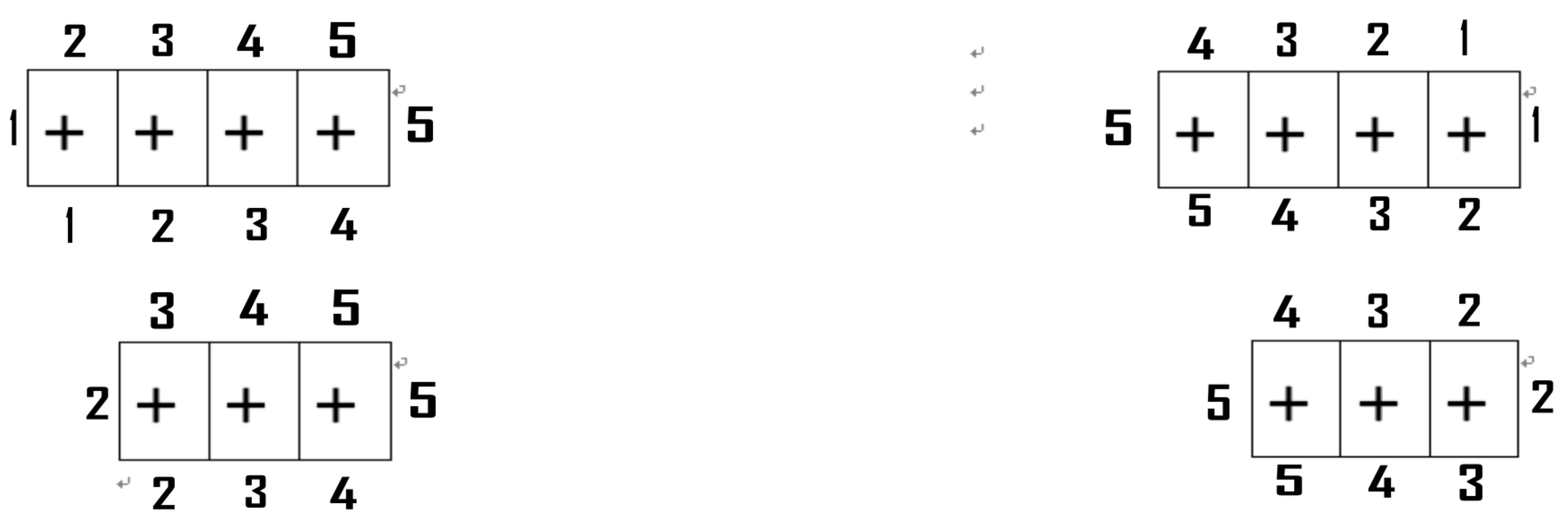}
\end{figure}

\begin{figure}[b]
	\includegraphics[width=1.2\textwidth]{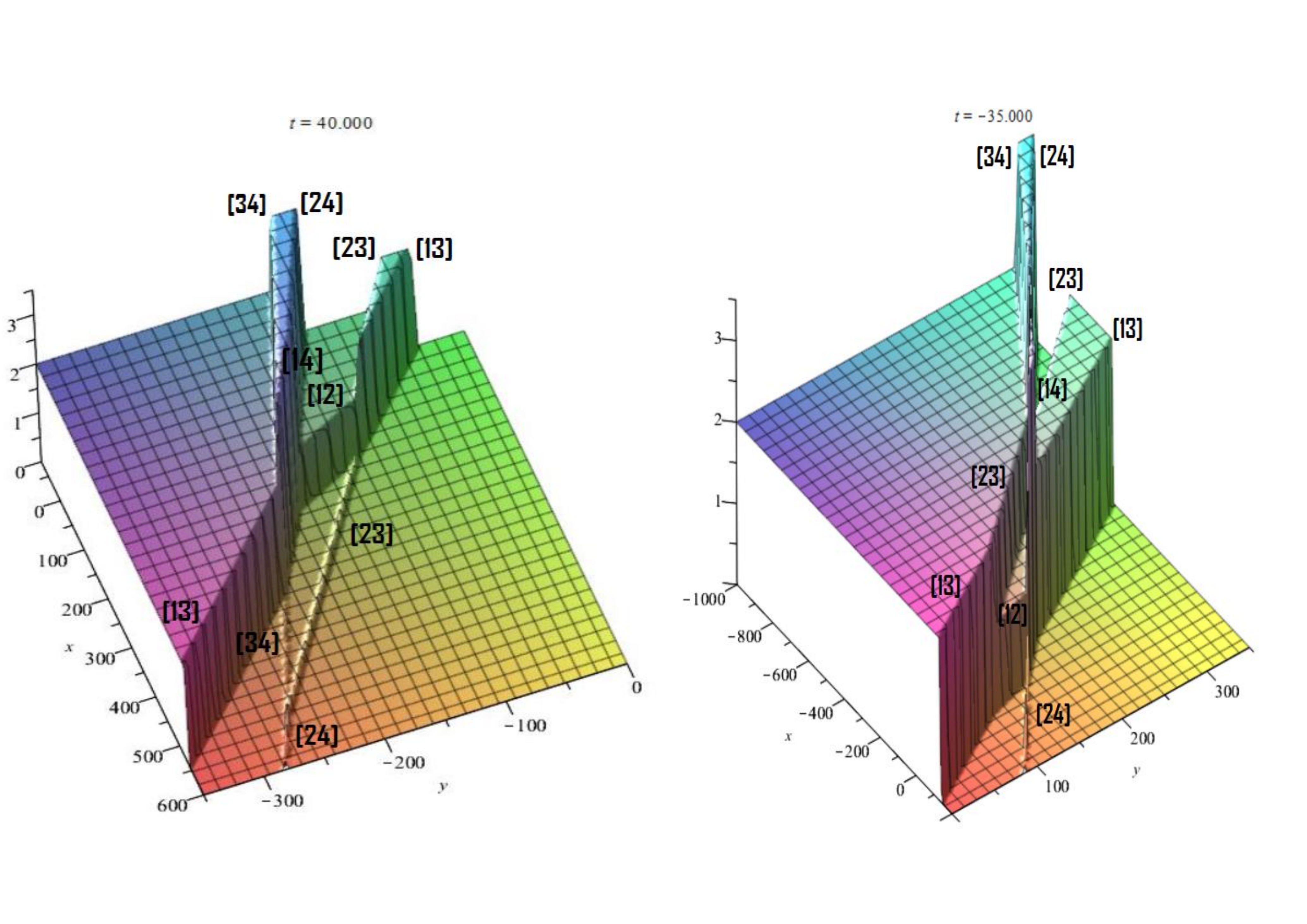}
	\caption{{$a = 10, b = 20, c = 30, d = 70,  k_1 = 0, k_2 = 1, k_3 = 2, k_4 = 3.5$ }}
\end{figure}

\begin{figure}[b]
		\includegraphics[width=1\textwidth]{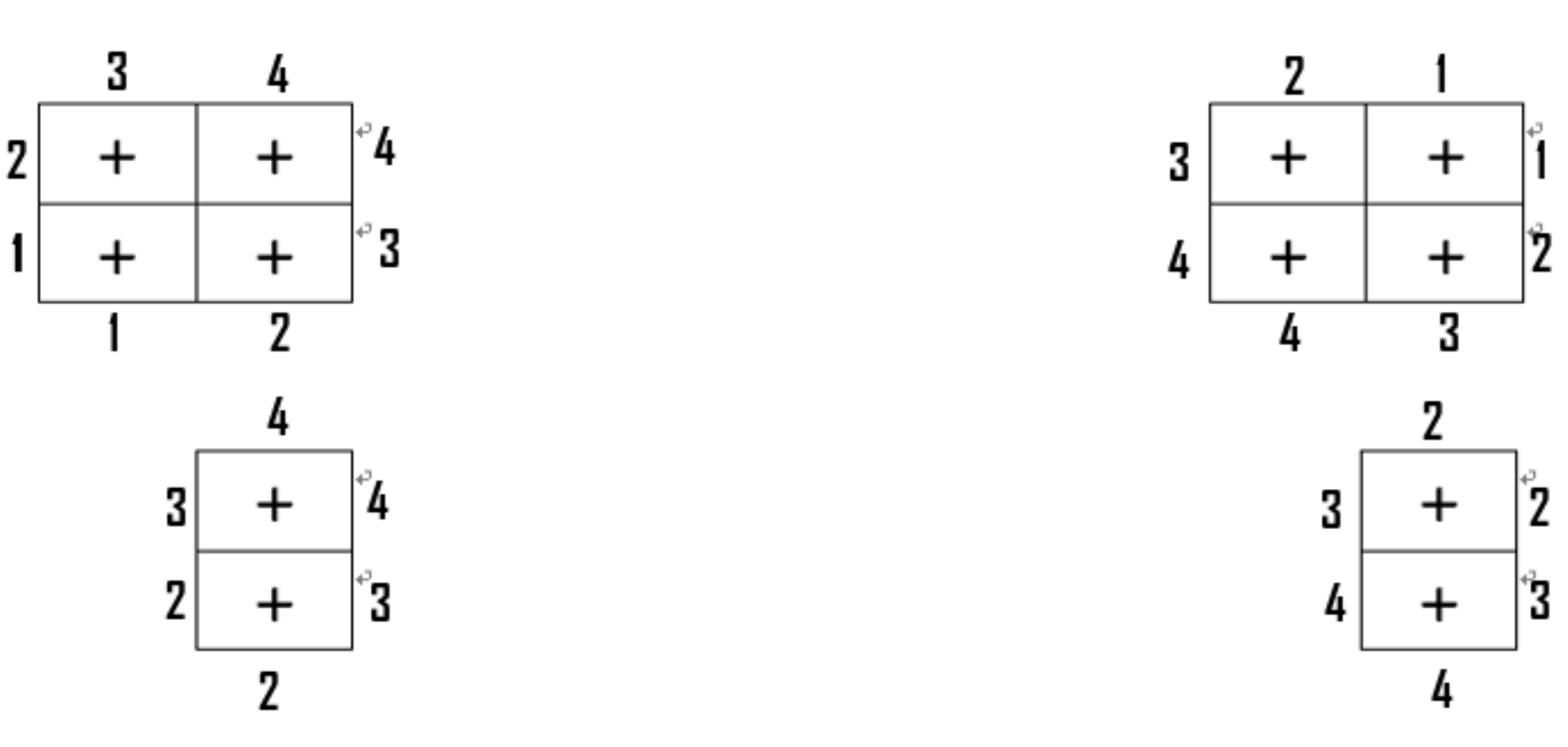}
	
\end{figure}

\begin{figure}[b]
	\includegraphics[width=1.1\textwidth]{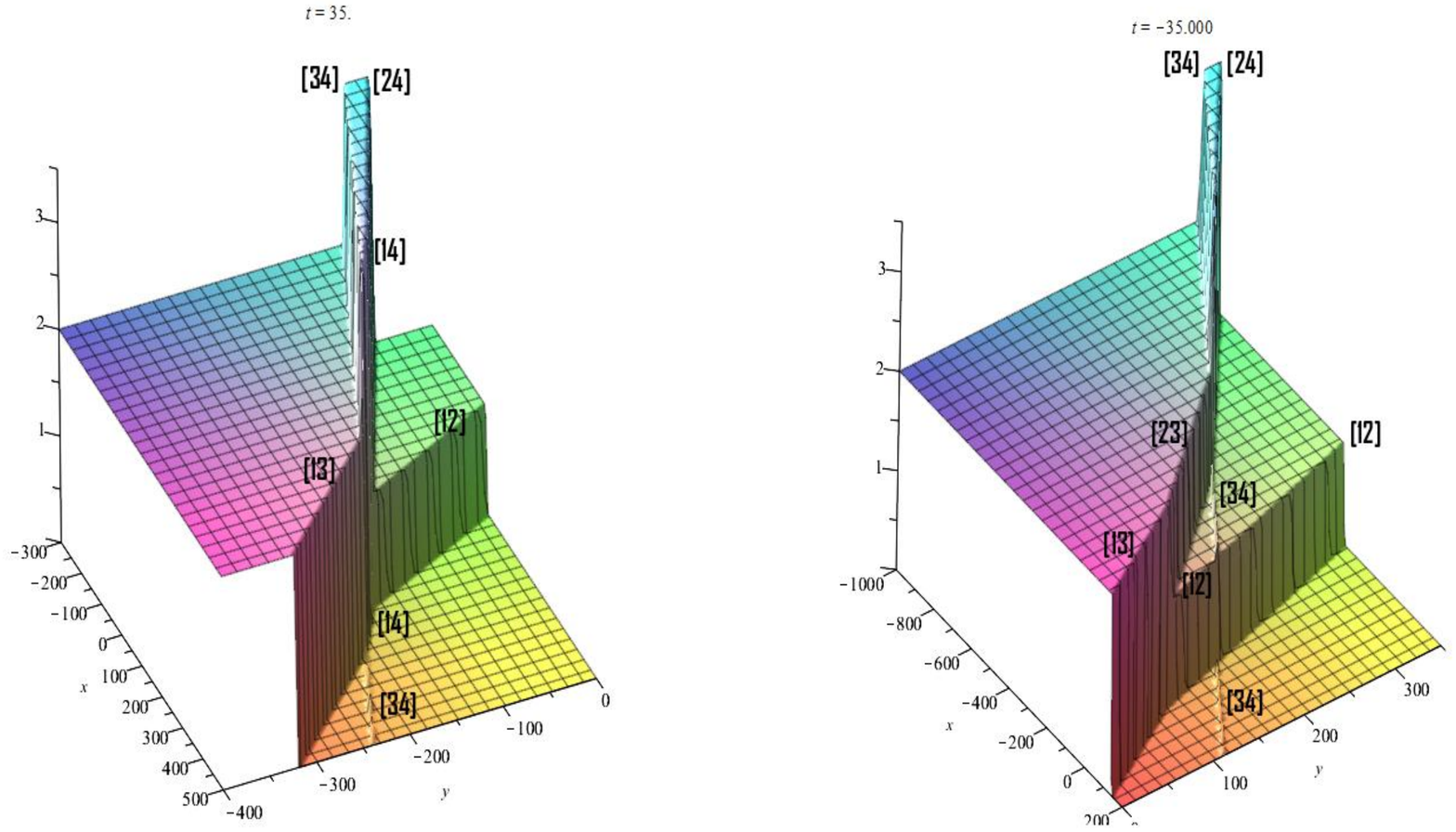}
	\caption{{$ a = 10, b = 20, c = 30, d = 60,  k_1 = 0, k_2 = 1, k_3 = 2, k_4 = 3.5 $}}
\end{figure}

\begin{figure}[b]
		\includegraphics[width=1\textwidth]{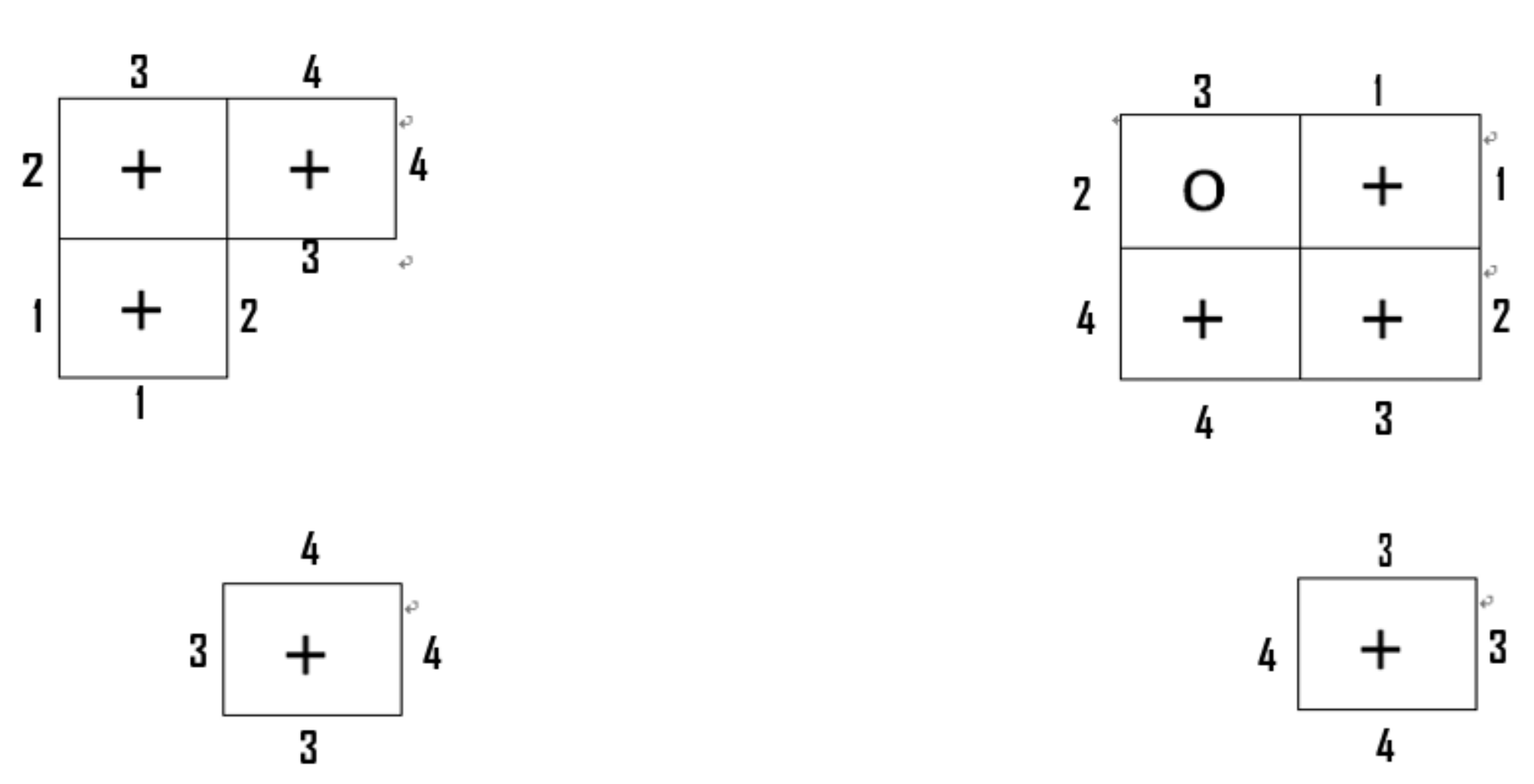}
	\end{figure}

\begin{figure}[b]
	\includegraphics[width=1.16\textwidth]{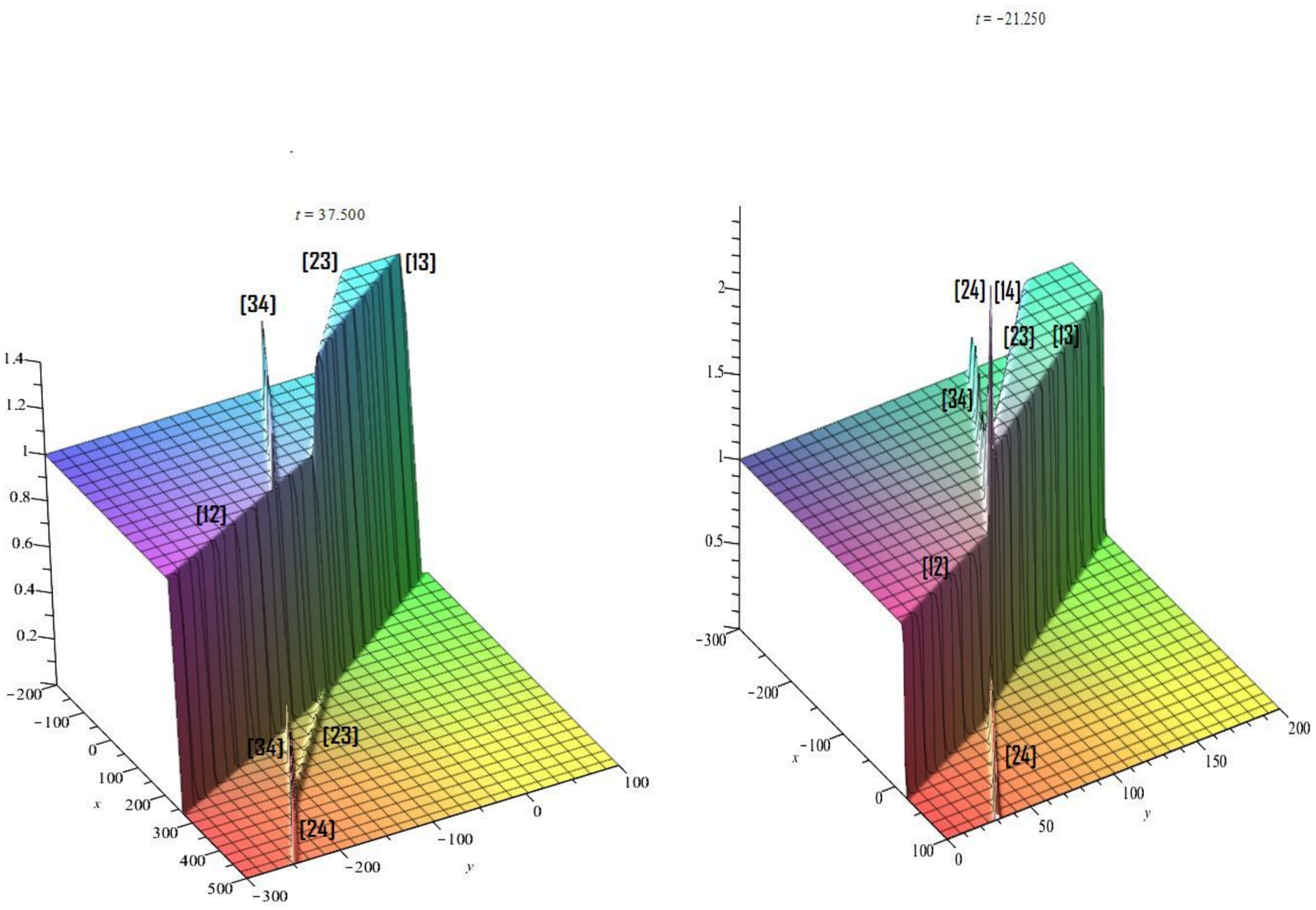}
	\caption{{ $ a = 100, b = 400, c = 500,  k_1 = 0, k_2 = 1, k_3 = 1.4, k_4 = 2.5 $}}
\end{figure}

\begin{figure}[b]
		\includegraphics[width=0.95\textwidth]{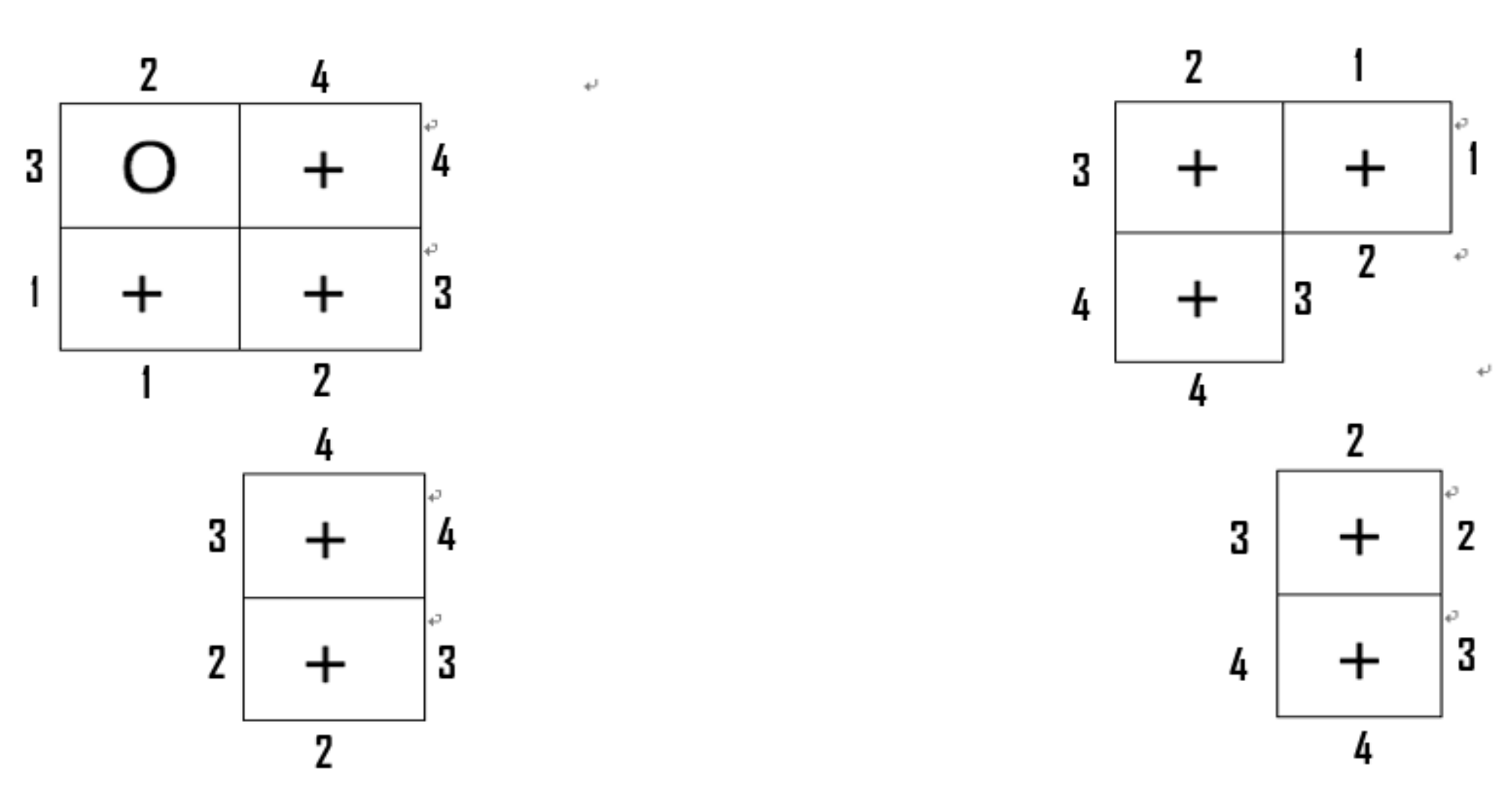}
	\label{fig:Five}
\end{figure}


\begin{thebibliography}{99}
\bibitem{bc}G. Biondini and S. Chakravarty, Soliton solutions of the Kadomtsev-Petviashvili II equation, J. Math. Phys. 47, 033514 (2006) 
\bibitem{ck} Sarbarish Chakravarty, Tim Lewkow and Ken-Ichi Maruno, On the construction of the KP line-solitons and their interactions, Applicable Analysis, Vol.89(2010), p.529-p.545
\bibitem{jh1} Jen-Hsu Chang, The Mach-type soliton in the Novikov-Veselov Equation, SIGMA 10 (2014), 111, 14 pages, 2014
\bibitem{jh2}Jen-Hsu Chang, The Interactions of Solitons in the Novikov-Veselov Equation, Applicable Analysis, vol. 95 (2016), p.1370-p.1388, arXiv:1310.4027 
\bibitem {jh3}Jen-Hsu Chang,  Soliton interaction in the modified Kadomtsev–Petviashvili-(II) equation,  To appear in Applicable Analysis, 2019, arXiv:1705.04827
%\bibitem{ds} G.C. Das, J. Sarma, Evolution of solitary waves in multicomponent plasmas, Chaos Soliton Fract. 9 (1998),  p.901-p.911
\bibitem{dk}V G Dubrovsky and B G Konopelchenko, On the interrelation between the solutions of the mKP and KP equations via the Miura transformation, Journal of Physics A: Mathematical and General, Vol. 24, N0. 18, 1991, p.4315-p.4324
\bibitem{hi} R. Hirota, The Direct Method in Soliton Theory, Cambridge Univ. Press, 2004
\bibitem{ko} S. Chakravarty and Y. Kodama, Soliton solutions of the KP equation and application to shallow water waves,
Stud. Appl. Math., 123 (2009),  p.83-p.151.
\bibitem{hw} Yu-tin Huanga and CongKao Wen, ABJM amplitudes and the positive orthogonal Grassmannian, Journal of High Energy Physics, 2 (2014), p.1-p.51
\bibitem{koo}Yuji Kodama, Young diagrams and N-soliton solutions of the KP equation, J. Phys. A 37 (2004), p.11169–p.11190 
\bibitem{ko1}Yuji Kodama, KP solitons in shallow water, J. Phys. A: Math. Theory 43, 434004 (2010) (54 pp), arXiv:1004.4607
\bibitem{ko2}S. Chakravarty and Y. Kodama, A generating function for the N-soliton solutions of the Kadomtsev-Petviashvili II equation,  Contemp. Math. 471, p.47-p.67 (2008) 
\bibitem{ko3}Y. Kodama and L. Williams, KP solitons and total positivity for the Grassmannian,  Invent. Math. 198, p.637–p.699 (2014) 
\bibitem{ko5}Y. Kodama and H. Yeh: The KP theory and Mach reﬂection,  J. Fluid Mech. 800, p.766–p.786 (2016)
\bibitem{ko6} Y. Kodama, KP Solitons and the Grassmannians:  Combinatorics and Geometry of Two-Dimensional Wave Patterns, Springer Briefs in Mathematical Physics, Springer Nature Singapore, 2017
\bibitem{kn} B.G. Konopelchenko, On the gauge-invariant description of the evolution equations integrable by Gelfand-Dikij spectral problems, Physics Letters A, Vol 92, Issue 7, 1982, p.323-p.327
\bibitem{kd} B. G. Konopelchenko and V. G. Dubrovsky, Inverse Spectral Transform for the Modified Kadomtsev-Petviashvili Equation, Volume 86, Issue 3, 1992, p. 219-p.268
\bibitem{kon} B. G. Konopelchenko, Solitons in Multi-Dimensions, World Scientific Publishing Company, Singapore, 1993
\bibitem{jm} M. Jimbo and T. Miwa, Solitons and Infinite Dimensional Lie Algebras, Publ RIMS, Kyoto Univ., 19 (1983), p.943-p.1001
\bibitem{lp}Jyh-Hao Lee and  Oktay K. Pashaev, Soliton Resonances for MKP-II, Theor.Math.Phys. 144 (2005) 995-1003; Teor.Mat.Fiz. 144 (2005), p.133-p.142 
\bibitem {os}  W. Oevel and W. Strampp, Constrained KP Hierarchy and Bi-Hamiltonian Structures, Commun. Math. Phys. 157 (1993), p.51-p.81 
\bibitem{po}Alexander Postnikov, Total Positivity, Grassmannian, and Networks, arXiv:math/0609764, 2006
\bibitem {br}Bo Ren, Interaction solutions for mKP equation with nonlocal symmetry reductions and CTE method, Physica Scripta, Vol. 90(2015), No.6, 065206
\bibitem{sp} A.V. Slunyaev, E.N. Pelinovsky, The role of multiple soliton interactions in generation of rogue waves: the mKdV framework, Phys. Rev. Lett., 117 (2016), 214501 
\bibitem {zz}Da-Jun Zhang , Song-Lin Zhao , Ying-Ying Sun and Jing Zhou, Solutions to the modiﬁed Korteweg–de Vries equation, Reviews in Mathematical Physics, Vol. 26, No. 7 (2014), 1430006
%\bibitem{sc}Zhi-Yuan Sun , Yi-Tian Gao , Xin Yu , Xiang-Hua Meng , Ying Liu, Inelastic interactions of the multiple-front waves for the modified Kadomtsev-Petviashvili equation in fluid dynamics, plasma physics and electrodynamics, Wave Motion, 46 (2009),  p.511-p.521
% \bibitem{vd} V. Veerakumar, M. Daniel, Modified Kadomtsev-Petviashvili (MKP) equation and electromagnetic soliton, Math. Comput. Simulat. 62 (2003), p.163-p.169.
% \bibitem{xz} T. Xu, H.Q. Zhang, Y.X. Zhang, J. Li, Q. Feng, B. Tian, Two types of generalized integrable decompositions and new solitary-wave solutions for the modified Kadomtsev-Petviashvili equation with symbolic computation, J. Math. Phys. 49 (2008) 013501.
\end{thebibliography}
\end{document}